\documentclass{emulateapj}

\newcommand{\kms}{km~s$^{-1}$}
\newcommand{\Ni}{$^{56}$Ni~}
\newcommand{\Co}{$^{56}$Co~}
\newcommand{\dmf}{$\Delta$M$_{15}$}
\usepackage{psfig}
\usepackage{amsmath}
\usepackage{amssymb}
\usepackage{graphicx}

\received{} 
\accepted{} 
\shortauthors{D.~Kasen {\it et al.}}
\shorttitle{Hole in SNe~Ia}

\begin{document}
\bibliographystyle{apj}

\title{Could There Be A Hole In Type Ia Supernovae?}

\author{Daniel~Kasen, Peter~Nugent, R.C.~Thomas \& Lifan~Wang}
\affil{Lawrence Berkeley National Laboratory, M.S. 50-F, 1
Cyclotron Road, Berkeley, CA 94720}

\email{e-mail: dnkasen@panisse.LBL.gov}

\begin{abstract}
In the favored progenitor scenario, Type Ia supernovae (SNe~Ia) arise
from a white dwarf accreting material from a non-degenerate companion
star.  Soon after the white dwarf explodes, the ejected supernova
material engulfs the companion star; two-dimensional hydrodynamical
simulations by \cite{Marietta} show that, in the interaction, the
companion star carves out a conical hole of opening angle
30$^\circ$-40$^\circ$ in the supernova ejecta.  In this paper we use
multi-dimensional Monte Carlo radiative transfer calculations to
explore the observable consequences of an ejecta-hole asymmetry.  We
calculate the variation of the spectrum, luminosity, and polarization
with viewing angle for the aspherical supernova near maximum light.
We find that the supernova looks normal from almost all viewing angles
\emph{except} when one looks almost directly down the hole.  In the
latter case, one sees into the deeper, hotter layers of ejecta.  The
supernova is relatively brighter and has a peculiar spectrum
characterized by more highly ionized species, weaker absorption
features, and lower absorption velocities.  The spectrum viewed down
the hole is comparable to the class of SN~1991T-like supernovae.  We
consider how the ejecta-hole asymmetry may explain the current
spectropolarimetric observations of SNe~Ia, and suggest a few
observational signatures of the geometry.  Finally, we discuss the
variety currently seen in observed SNe~Ia and how an ejecta-hole
asymmetry may fit in as one of several possible sources of diversity.
\end{abstract}

\section{Introduction}

\subsection{Asymmetry of Type Ia Supernovae}
Some Type~Ia supernovae (SNe~Ia) are known to be aspherical; direct
evidence for this comes from optical spectropolarimetric observations.
Because a spherically symmetric system has no preferred direction, the
polarization integrated over the projected supernova surface cancels
-- detection of a non-zero intrinsic polarization demands some degree
of asymmetry.  The measured intrinsic polarization of SNe~Ia is
relatively small, but certainly detected in a few cases.  Pre-maximum
observations of the normal Type~Ia SN~2001el using the ESO Very Large
Telescope found an intrinsic polarization level of $\sim0.3\%$, which
decreased at later epochs \citep{Wang_pol_01el, Kasen_01el}.
Intrinsic polarization of $\sim 0.7\%$ was also measured for the
underluminous and spectroscopically peculiar SN~1999by
\citep{Howell_99by}.

The geometry of SNe~Ia must be closely tied to the supernova explosion
physics and progenitor system, both of which are still under debate.
But little is known about the shape of the ejecta.  For both SN~1999by
and SN~2001el we do know that the bulk of the ejecta obeyed a nearly
axial symmetry.  This is because in both cases, after subtraction of
the interstellar polarization, the polarization angle was fairly
constant over the entire spectral range (with the exception of an
unusual high-velocity calcium feature in SN~2001el).  Most theoretical
attempts at modeling the spectropolarimetry have so far assumed the
ejecta was ellipsoidal
\citep{Howell_99by,Wang_96x,Jeffery_87a,Hoeflich_91}.  A shape like
this might arise, for example, in the explosion of a rapidly rotating
progenitor star.

Another potential cause of asymmetry in SNe~Ia is the binary nature of
the progenitor system.  In the favored progenitor scenario (the
\emph{single-degenerate scenario}; see \cite{Branch_progenitor} and
references therein), SNe~Ia arise from a white dwarf accreting
material from a non-degenerate companion star.  The companion may be
either a main sequence star, a red-giant, or a subgiant; as it is
close enough to be in Roche-lobe overflow, it subtends a substantial
solid angle from the perspective of the white dwarf.  The supernova
explosion occurs when the white dwarf has accreted enough matter that
the densities and temperatures at the center are sufficient to ignite
carbon, just below the Chandrasekhar limit.  The ejected supernova
material moves at a few percent of the speed of light and soon after
the explosion (from minutes to hours) engulfs the companion star.  In
the impact it would not be surprising if a substantial asymmetry was
imprinted on the supernova ejecta.

The ejecta-companion interaction has been studied with two-dimensional
hydrodynamical models by \cite{Fryxell_Arnett}, \cite{Livne_binary},
and most recently and extensively by \cite{Marietta}.  These studies
were primarily concerned with the fate of the companion star, in
particular how much hydrogen gets stripped from its outer envelope.
Stripped hydrogen may appear as narrow Balmer emission lines in the
supernova spectrum, which if observed might provide direct evidence of
a binary progenitor system.  With the advance of spectropolarimetric
observations, however, the nature of SN~Ia asphericity becomes another
relevant test of the single-degenerate progenitor scenario.  In their
hydrodynamical models, \cite{Marietta} find that the impact with the
companion star carves out a conical hole in the supernova ejecta.  The
opening angle of the hole is 30$^\circ$-40$^\circ$, and because the
ejecta is moving supersonically, the hole does not close with time.
The final configuration is axially symmetric, as was seen in the
polarization observations of SN~2001el.

In this paper we use multi-dimensional radiative transfer calculations
to address the possibility of SNe~Ia having an ejecta hole asymmetry.
We calculate the variation of the spectrum, luminosity, and
polarization with viewing angle for the aspherical supernova near
maximum light.  In contrast to the ellipsoidal models, the angular
variations in an ejecta-hole geometry can be rather extreme,
especially when one looks near the hole itself.  These variations
would necessarily introduce some diversity into the observed
properties of SNe~Ia.  The question is: exactly what sort of diversity
arises in the ejecta-hole geometry, and does this fit in with the
diversity already known to exist in SNe~Ia?

While SNe~Ia are considered to be a rather homogeneous class of
objects, they do show some variety in their spectral and photometric
properties.  The observed peak magnitudes of SNe~Ia vary by $\sim 0.3$
mag, and the brightness is found to correlate with the width of the
light curve \citep{Phillips93}.  The spectra of SNe~Ia can be
classified as either normal or peculiar \citep{Branch_normal}.  The
peculiar spectra have feature strengths at maximum light that differ
from ``normal'' cases (such as SN~1981B), and are usually subdivided
into two classes: SN~1991bg-like supernovae have a broad \ion{Ti}{2}
absorption trough not seen in the normals \citep{Filippenko_91bg};
SN~1991T-like supernovae have weak or absent features from singly
ionized species but noticeable \ion{Fe}{3} lines
\citep{Filippenko_91T,Phillips_91T,Jeffery_92}.  Not all supernovae
fit cleanly into the classification scheme.  In its pre-maximum
spectra, SN~1999aa resembled SN~1991T, but by maximum light it had
begun to look much more normal, with \ion{Si}{2} and \ion{Ca}{2} lines
that were stronger than SN~1991T but weaker than normal
\citep{Li_peculiar}.  As such SN~1999aa is considered by some to be an
intermediate link between the normal and the SN~1991T-like supernovae.
Other observations have uncovered singular objects like SN~2000cx
\citep{Li_00cx} and SN~2002cx \citep{Li_02cx}, that while resembling
SN~1991T in some ways (weak \ion{Si}{2}, strong \ion{Fe}{3} lines)
showed other peculiarities that were unique.  Additional spectral
diversities include the abnormally high photospheric velocities of
SN~1984A \citep{Branch_84A} and the detached, high velocity features
seen in several supernovae
\citep{Hatano_94d,Wang_pol_01el,Thomas_00cx}.  The diversity of SNe~Ia
is thus multi-faceted, a point we return to in the conclusion.

\section{The Ejecta-Hole Model}

\subsection{Density and Composition Structure}

\begin{figure}
\begin{center}
\psfig{file=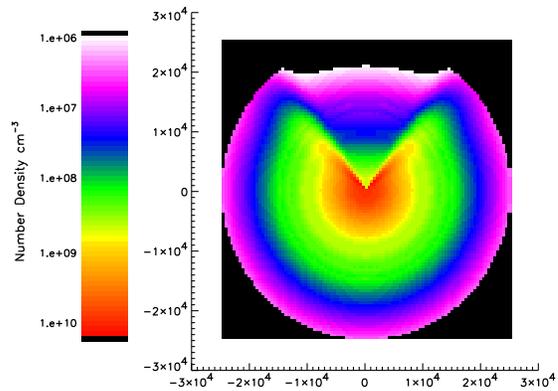,width=3.3in}
\caption{Density structure of the ejecta-hole model near maximum
light (20 days after explosion).
\label{dens_plot}}
\end{center}
\end{figure}

The ejecta model used in the calculations is based on the spherical W7
explosion model \citep{Nomoto_w7}, which has often been used in
spherical radiative transfer calculations to model the spectra of
normal SNe~Ia \citep{Lentz_94d, Jeffery_92, Nugent_hydro}.  The
composition structure of W7 consists of an inner \Ni zone ($3000 < v <
9000$ \kms), a middle zone of intermediate mass elements (9000 \kms $<
v <$ 15,000 \kms) and an outer unburned region of carbon-oxygen rich
material ($v > 15,000$ \kms).  In our calculations we found it
necessary to make one adjustment to the compositions; to reproduce
the depth and width of the \ion{Ca}{2} H\&K feature in a normal SN~Ia,
we needed to increase the calcium abundance by a factor of 10 in the
outer C-O region.  The difficulty W7 has in fitting the \ion{Ca}{2}
H\&K feature has already been noted by \cite{Lentz_94d} in the context
of detailed NLTE models.  The lack of burned material above 15,000
\kms~may indicate a weakness of the parameterized deflagration
explosion model used.

To introduce an ejecta hole into the spherical model, we describe the
density structure by an analytic function that in the radial direction
well reproduces W7:
\begin{equation}
\rho(v,\theta) = \rho_0 \exp(-v/v_e) F(\theta)  
\end{equation}
where $v_e = 2,500$ \kms~and $\rho_0$ is set by the condition that
total mass of the ejecta equals a Chandrasekhar mass.  $F(\theta)$ is
an angular density variation function which would equal one in a
spherical model.  For the ejecta-hole model, we use a constructed
function which resembles the structure seen in the interaction models
of \cite{Marietta}.  The conical hole has a half opening angle of
$\theta_H = 40^\circ$ and the density in the hole is a factor $f_H =0.05$
less then that outside.  The material that is displaced from the hole
gets piled up into a density peak just outside the hole edge, with
angular size $\theta_P = 20^\circ$.  The function invented to reproduce these 
features is:
\begin{equation}
F(\theta)  = 
f_H + (1-f_H) \biggl(\frac{x^n}{1 + x^n}\biggr)
\biggl(1 + A e^{-(\frac{\mu-\mu_H}{\mu_P})^2} \biggr)
\label{mu_func}
\end{equation}
where
\begin{equation}
x = \frac{1 - \mu}{\mu_H}
\end{equation}
where $\mu=\cos\theta$ and $n=8$.  The constant $A$ is set by the
condition that the mass within a shell is equal to that in the
spherical model (i.e the integral of $F(\theta)$ over solid angle is
equal to $4\pi$).  The density structure is shown in
Figure~\ref{dens_plot}.

This analytic function does not capture all the complexity present in
a hydrodynamical model; for example, \cite{Marietta} point out that
the opening angle of the hole is slightly smaller at high velocities
than low velocities ($\sim 30^\circ-35^\circ$ as opposed to
$40^\circ$).  Of course, the benefit of using a simple analytic
function is that it isolates the essential geometrical consequences of
a hole asymmetry; in addition it allows us to test in a parameterized
way how varying the ejecta hole structure affects the observable
signatures.  Once the general ideas are understood, one can perform
more specific calculations using hydrodynamical models spanning a wide
range of initial progenitor conditions.

In the ejecta/companion interaction, as much as 0.1-0.5 $M_\sun$ of
hydrogen rich material can be stripped and ejected from the companion
star \citep{Wheeler_strip,Marietta}.  This material is not included in
our calculations.  The vast majority of the stripped material has low
velocity ($v < 1000$ \kms) and sits at the center of the ejecta, where
it will not affect the spectrum or polarization near maximum light.  A
small amount of material may be ejected at high velocities and could
be related to the high-velocity spectral features seen, for example,
in SN~2001el and SN~2000cx.  Both \cite{Branch_00cx} and
\cite{Thomas_00cx} have suggested an identification of high-velocity
$H_\beta$ in SN~2000cx, which if correct would strongly suggest that
the material was associated with the companion in some way.  While not
addressed in this paper, the observable consequences of the stripped
material should be explored further with multi-dimensional transfer
calculations that include a NLTE treatment of hydrogen.

\subsection{Monte Carlo Code}
Our calculations are carried out with a Monte-Carlo (MC) radiative
transfer code, described in detail in \cite{Kasen_MC}.  The code
applies principles described in, e.g.
\cite{Lucy_Radeq,Mazzali_MC,Code-blobs}.  In the MC approach, photon
packets are emitted from within the supernova envelope and tracked
through randomized scatterings and absorptions until they escape the
atmosphere.  Each packet is of a specific wavelength and contains a
Stokes vector which describes its polarization state.  All packets
escaping in a certain direction are collected to construct the
spectrum and polarization of the supernova from that viewing angle.
Our calculations use 100 angular bins, equally spaced in $\cos\theta$,
to collect escaping photon packets.  While the code can handle
arbitrary three-dimensional (3-D) geometries, for the
axially-symmetric models of this paper we use a two-dimensional (2-D)
Cartesian grid of $10^4$ cells to represent the supernova atmosphere.

One important issue in multidimensional MC transfer is where to place
the emission source of photon packets.  While most MC calculations
emit packets from a spherical inner boundary surface (the inner `light
bulb''), in an ejecta-hole model such an approach would provide a poor
representation of the geometry.  Therefore we have developed an
integrated multi-dimensional gamma ray transfer MC to determine
exactly where radioactive energy from decaying \Ni and \Co is
deposited in the supernova envelope (see \S~\ref{gamma_ray}).  The
optical photon packets are then emitted from individual cells
throughout the atmosphere, proportional to the local instantaneous
energy deposition rate.  There is no inner boundary surface, and
photons are allowed to propagate throughout the entire supernova
envelope, including the optically thick center.  Overall, this
approach is likely a good approximation to the actual conditions in
SNe~Ia, as the luminosity at maximum light is dominated by radioactive
energy deposition.  However a proper treatment would also take into
account diffusive energy stored in the supernova envelope by solving
the full time-dependent radiative-hydrodynamics problem.

The opacities used in the calculation are electron scattering and
bound-bound transitions; we ignore bound-free and free-free opacities
as these are much less important in SN~Ia atmospheres
\citep{Pinto-Eastman_II}.  Excitation and ionization are computed
assuming LTE, where the temperature structure of the atmosphere is
determined self-consistently using an iterative approach which imposes
radiative-equilibrium.  Line processes included are absorption and
scattering, according to a two level atom with thermalization
parameter $\epsilon= 0.05$ \citep{Nugent_hydro}.  Because the
detailed NLTE source function of the material is not calculated,
packets are initially emitted according to a blackbody distribution
with characteristic temperature $T_{bb}$.  We choose $T_{bb}$ so as to
reproduce the continuum in the red end of the observed spectrum; the
blue end of the spectrum shows very little dependence on $T_{bb}$, as
packets with $\lambda \la 6000~\AA$ are absorbed and re-emitted in
lines.  The photon packets are initially emitted unpolarized but
acquire polarization by electron scattering.  Line scattered light is
assumed to be unpolarized due to complete redistribution, as in the
models of \cite{Hoflich-93J}, \cite{Howell_99by}, and
\cite{Kasen_01el}.

\section{Results}
We have computed the gamma-ray deposition, optical spectrum, relative
luminosity and polarization of the ejecta hole model near maximum
light (20 days after the explosion) as a function of the viewing angle
$\theta$.  Because the current Monte Carlo code is not time-dependent,
we leave for future work the effect on the asymmetry on the light
curve.  For the maximum light model, the total luminosity used is $L =
1.4\times10^{43}$ ergs and the emission temperature $T_{bb} = 11,000$
K.  We discuss the various results in turn.

\subsection{Gamma Ray Deposition} \label{gamma_ray}
In the W7 explosion models, $\sim 0.6 M_\sun$ of radioactive \Ni is
synthesized and will power the supernova luminosity.  The majority of
the decay energy from \Ni and its daughter \Co is released as gamma
rays, which deposit their energy in the supernova ejecta primarily
through Compton scattering.  It takes only a few Compton scatterings
for a gamma ray to give up the majority of its energy to fast
electrons, which are in turn assumed to be thermalized locally.  We
compute the gamma ray energy deposition with a MC transfer routine
that includes Compton and photo-electric opacities and also produces
gamma ray spectra.

In a spherical SN~Ia model, the gamma ray trapping is very effective
at maximum light.  In the inner \Ni zone, the mean free path to
Compton scattering is only $\sim$300 \kms~and so gamma rays deposit
energy nearly coincident to where they are created; only about 4\% of
the gamma ray energy escapes the atmosphere.  Inside an ejecta hole,
on the other hand, the mean free path is 20 times greater due to the
lower density.  Gamma rays generated in the hole can therefore escape
the atmosphere, at least those that are emitted in the outward
direction.  This energy loss is not very significant, however, as the
hole is largely evacuated and contains less then 1\% of the total \Ni
mass.  The material that has been displaced from the hole (containing
$\sim$11\% of the total \Ni mass) is piled up around the hole edge,
where the density is high, and the gamma ray trapping is even more
efficient than in a spherical model.  Thus we find the perhaps
unexpected result that the ejecta hole actually slightly enhances the
gamma ray trapping at maximum light, from 96\% to 97\%.

Using Arnett's law as a rough rule of thumb \citep{Arnett_typeI}, the
luminosity of a SN~Ia at maximum light should be comparable to the
instantaneous rate of energy deposition.  One therefore expects that
in the ejecta-hole model the total luminosity at peak will be close to
(perhaps slightly greater than) a spherical model.  In other words,
although the aspherical supernova will appear significantly dimmer or
brighter depending upon the viewing angle (as we will see in
\S~\ref{peak_mags}), the specific luminosity integrated over
\emph{all} viewing angles will not be entirely different from the
spherical case.  However, time-dependent calculations are needed to
properly address this question, and so we leave it for future work.

\subsection{The P-Cygni Profile}

\begin{figure}
\begin{center}
\psfig{file=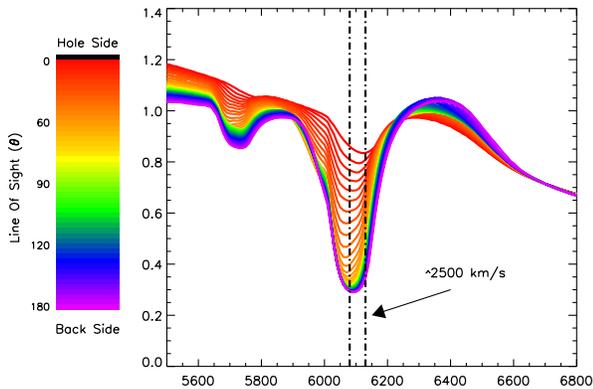,width=3.3in}
\caption{Flux profile of the \ion{Si}{2} 6150 line (at maximum light)
from various viewing angles (the view down the hole is the top-most
spectrum). When viewed down the hole ($\theta = 0^\circ$) the
absorption trough is weaker and has a lower velocity by $\sim 2,500$
\kms.  Silicon is the only species included in this calculation.
\label{line_los}}
\end{center}
\end{figure}

Line opacity in a spherical, expanding SN atmosphere gives rise to the
well known P-Cygni profile -- i.e. a blueshifted absorption trough
with a redshifted emission peak.  An ejecta-hole asymmetry
dramatically alters the line profile from some lines of sight, as
shown in Figure~\ref{line_los}.  The major effects are readily
apparent: in the typical P-Cygni formation, material in front of the
photosphere obscures the light below and gives rise to the blueshifted
absorption feature.  When one looks down the ejecta hole ($\theta <
\theta_H$), the density of this obscuring material is much lower and
the line absorption features are thus much weaker.  There is little
change, however, in the redshifted emission component.
\cite{Rollin_clumpy} have pointed out that asymmetries have the most
dramatic effect on absorption features, as the absorption depth is
related directly to how much of the photosphere is covered by line
opacity.

As one looks away from the hole, the line absorption depth increases
rapidly, until for $\theta > \theta_H$ the depth is equal to that of
the spherical model.  For side-on views $(\theta \approx 90^\circ$),
the hole is in the emission region -- because some emitting material
is then lacking one expects the P-Cygni emission feature to be
depressed near the line wavelength center.  The missing material,
however, amounts to only 11\% of the total emitting area, so the
effect is hardly noticeable.  For $\theta > \theta_H$, the line
profile changes very little with viewing angle.

The minima of the absorption features are also less blueshifted when
viewed down the hole, by about 2000-3000 \kms.  This is because the
hole allows one to see relatively deeper into the ejecta. In a
spherical model, P-Cygni features are formed primarily by material at
or above the supernova photosphere, while layers below will not be
visible until the expanding supernova thins out and the photosphere
recedes.  For views down the ejecta hole, however, the electron
scattering photosphere has an odd shape, resembling the conical hole
of Figure~\ref{dens_plot}.  As radiation streams radially out the
hole, absorption features are caused by relatively deeper layers of
ejecta.  This deeper material will tend to be hotter, more ionized and
perhaps of a different composition than the material in the outer
layers.  One therefore expects that the features of more highly
ionized species will be relatively more prominent when the supernova
is viewed down the hole.  The exact line strengths depend, of course,
upon the temperature and ionization structure in the 2-D atmosphere,
which is calculated self-consistently in LTE in our models.

\subsection{Spectrum Near Maximum Light} \label{max_spectrum}

\begin{figure}
\begin{center}
\psfig{file=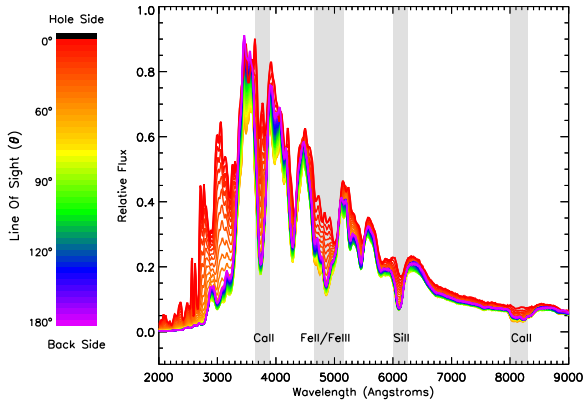,width=3.3in}
\caption{Spectrum of the ejecta-hole model near maximum light from
various viewing angles (the view down the hole is the top-most
spectrum).  Some important line features are highlighted.
\label{spec_los}}
\end{center}
\end{figure}

In sum, the spectrum in the ejecta-hole model will look the same as in
a spherical model for all lines of sight \emph{except} when one looks
almost directly down the hole ($\theta < \theta_H$).  In the latter
case, one sees a peculiar spectrum characterized by more highly ionized
species, weaker absorption features, and lower absorption velocities.
We show the variation of the maximum light spectrum with viewing angle
in Figure~\ref{spec_los}.  Notice in particular the dramatic effect
the hole has on the \ion{Si}{2} and \ion{Ca}{2} features, the iron
blend near 5000~\AA, and the UV region of the spectrum ($\lambda <
3500$~\AA).

\begin{figure}
\begin{center}
\psfig{file=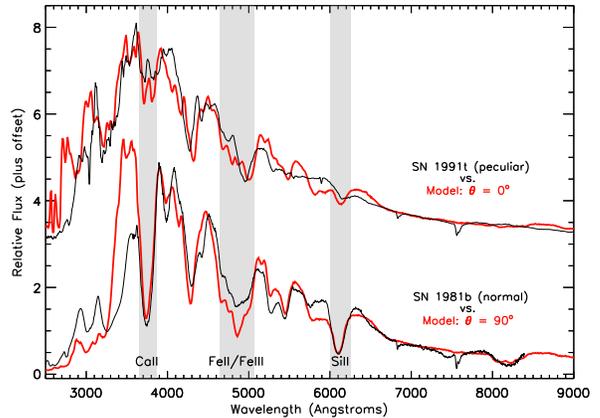,width=3.3in}
\caption{The maximum light spectra of the ejecta-hole model (thick red
lines) from two different viewing angles are compared to two observed
SNe~Ia (thin black lines).  Bottom: the view from the side
($\theta=90^\circ$) compared to the normal SN~1981B.  Top: the view
down the hole ($\theta = 0^\circ$) compared to the peculiar SN~1991T.
\label{max_compare}}
\end{center}
\end{figure}

Figure~\ref{max_compare} compares the model spectra to two well known
SNe~Ia.  The view away from the hole ($\theta=90^\circ$) resembles the
normal Type~Ia SN~1981B.  The model reproduces most of the major
spectral features, although there are a few discrepancies.  The most
obvious is that the flux peak near $3500~\AA$ is much to large in the
model.  Because the opacity at this wavelength is largely due to
\ion{Co}{2} lines, models which mix some \Ni out to higher velocities
can suppress the peak (see \cite{Branch_81b, Jeffery_92}).  The poor
match is also likely in a part due to the approximate treatment of
wavelength redistribution in our calculations (a constant
$\epsilon=0.05$, two level atom).

The spectrum down the hole ($\theta=0^\circ$) is clearly very
different than a normal SN~Ia.  We compare it to the peculiar
SN~1991T, which it resembles in the following ways: (1) the
\ion{Si}{2} absorption near 6150~\AA~is weak and has an unusually low
velocity ($v \approx 10,000$ \kms); in addition, the \ion{Si}{2}
absorption at 4000~\AA~is absent.  (2) The \ion{Ca}{2} H\&K feature is
weak and shows a ``split'' into two lines (due to \ion{Ca}{2} H\&K and
\ion{Si}{2} $\lambda3858$; \cite{Nugent_hydro}); in addition, the
\ion{Ca}{2} IR triplet absorption is absent.  (3) In the iron blend
near 5000~\AA, the broad \ion{Fe}{2} absorption is weak while the
sharper \ion{Fe}{3} feature to the red is prominent. (4) The
ultraviolet portion of the spectrum (2500~\AA $<\lambda< 3500$~\AA) is
much brighter down the hole, due to the decreased line blocking.

For now, the comparison of Figure~\ref{max_compare} is meant only to
illustrate that the spectrum emanating from the hole would be
categorized as having so-called SN~1991T-like peculiarities.  What
connection, if any, the hole asymmetry may have to SN~1991T itself
will be discussed further in the conclusion.  Note that there are also
apparent differences between SN~1991T and the model, among them: (1)
The \ion{S}{2} ``W-feature'' near 5500~\AA~is weak but visible in the
model, whereas no clear feature is seen in SN~1991T; (2) The model has
too much emission in the \ion{Si}{2} 6150 and \ion{Ca}{2} IR triplet
features. (3) The velocities of the \ion{Fe}{3} lines are too low in
the model, by about 2000 \kms.  The \ion{Fe}{3} lines are forming just
at the edge of the exposed iron/nickel core, so an explosion model
that had a slightly larger \Ni zone than W7 might provide a better
match SN~1991T.

As can be seen in Figure~\ref{spec_los}, the spectrum changes
continuously from peculiar to normal as the viewing angle is increased
from zero.  Some degree of peculiarity is seen for $\theta <
\theta_H$, but the further the viewing angle is from $0^\circ$, the
less intense the peculiarities.  For a viewing angle of $\theta_H
\approx 30^\circ$, for instance, the depths of the \ion{Si}{2} and
\ion{Ca}{2} features are about half that of the normal case, and the
iron blend near 5000~\AA~is dominated by \ion{Fe}{2} rather than
\ion{Fe}{3}. One might rather compare the model from this viewing
angle to SN~1999aa, which near maximum light was in many ways
intermediate between SN~1991T and a normal SN~Ia.

We have also experimented with varying the density structure of the
ejecta hole.  As can be expected, increasing the density in the hole
or decreasing the hole opening angle tames the asymmetry and produces
spectra with less intense peculiarities.  If the hole opening angle is
reduced below $\theta_H \la 20^\circ$ the spectral peculiarities are
minor from all lines of sight.  In the hydrodynamical models of
\cite{Marietta}, the hole opening angle is $40^\circ$ in the low
velocity layers, and $30^\circ-35^\circ$ in the outer high-velocity
layers, depending upon the nature of the companion star.  The hole
used in Figures~\ref{spec_los} and \ref{max_compare} ($\theta_H =
40^\circ$ in all layers) thus represents the extreme end of what one
might expect from their calculations.

\subsection{Peak Magnitudes}
\label{peak_mags}

\begin{figure}
\begin{center}
\psfig{file=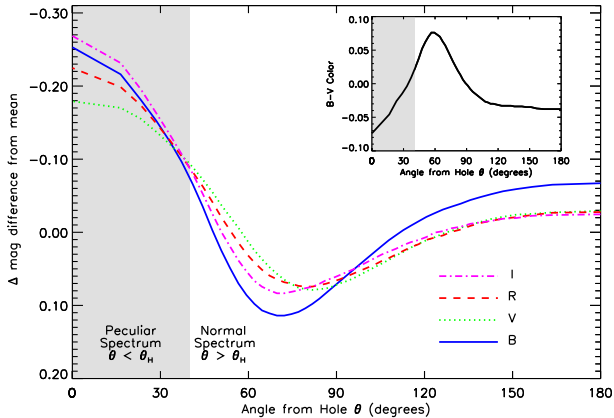,width=3.3in}
\caption{Variation with viewing angle of the $B$, $V$, $R$ and
$I$-band magnitudes of the ejecta-hole model near maximum light.  The
magnitudes are plotted relative to the mean magnitude averaged over
all viewing angles.  The inset shows the variation of the $B$-$V$
color.
\label{mags}}
\end{center}
\end{figure}

In the ejecta-hole model, the observed luminosity depends upon the
viewing angle (Figure~\ref{mags}).  When viewed down the hole, the
supernova is brighter by up to 0.25 mag in $B$.  This is because
photons more readily escape out the hole due to the lower opacities.
On the other hand, the supernova is dimmer than average when viewed
from the side ($\theta \approx 90^\circ$) because from this angle the
supernova is lacking a ``wedge'' of scattering material (see
Figure~\ref{block_plot}a).  Radiation that would normally have been
scattered into the $90^\circ$ view now flows straight out the hole and
goes into making the view down the hole brighter.

It is widely believed that observed SN~1991T-like supernovae are in
general overluminous, although the degree and regularity of this
overluminosity can be questioned \citep{Saha_91T}.  While
Figure~\ref{mags} suggests a similar relationship, keep in mind that
the total luminosity is a fixed parameter in this calculation -- the
figure only shows how this fixed luminosity gets distributed among the
various viewing angles.  In general, one expects the total luminosity
to depend predominately on the amount of \Ni synthesized in the
explosion, which will vary from supernova to supernova.  If a certain
SN~Ia has a very small \Ni mass, for example, then although the view
down the hole is the brightest of all possible viewing angles, the
supernova would still appear underluminous compared to a SN~Ia with
normal \Ni production.

The total dispersion about the mean in the ejecta hole model is $\sim
0.1$ mag in $V$ and $R$, and somewhat larger in $B$ ($\sim0.2$ mag) as a
result of the $B$-band's greater sensitivity to line opacity.  The
observed dispersion in SNe~Ia peak magnitudes is around 0.3 mag in the
$B$-band, and the brightness is found to correlate with the width of
the light curve \citep{Phillips93}.  These variations are believed to
be largely the result of varying amounts of \Ni synthesized in the
explosion.  After correction for the width-luminosity relation and
dust extinction (using the $B$-$V$ color), the observed dispersion is
reduced to $\sim 0.15-0.2$ mag \citep{Hamuy_96b}.  Some of this
so-called intrinsic dispersion is likely due to an asymmetry of some
sort; Figure~\ref{mags} suggests that in the particular case of an
ejecta-hole geometry, the asymmetry may in fact be the dominant
effect.  Note, however, that in the model the $B$-$V$ color roughly
correlates with peak magnitude -- thus correcting for dust extinction
with a $B$-$V$ color will tend to correct for the asymmetry also.  The
angular variation of the luminosity is also sensitive to the details
of the hole structure -- decreasing the hole size to $\theta_H =
30^\circ$, for example, decreases the $B$-band dispersion to $\sim$0.1
mag.

\subsection{Continuum Polarization}

\begin{figure}
\begin{center}
\psfig{file=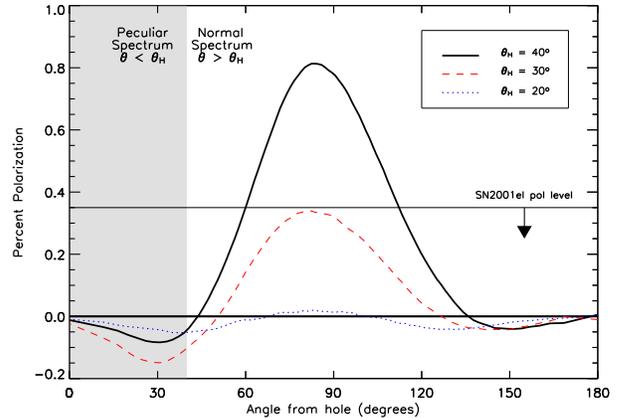,width=3.3in}
\caption{Continuum polarization of the ejecta-hole model near maximum
light as a function of viewing angle.  The solid black line is the
model used throughout the paper, while the red dashed and blue dotted
lines are models where the hole opening angle has been reduced to
$30^\circ$ and $20^\circ$ respectively.
\label{pol_cont}}
\end{center}
\end{figure}

The polarization is the most direct indication of asymmetry in the
ejecta.  Because a spherically symmetric atmosphere has no preferred
direction, the polarization integrated over the projected supernova
surface must cancel.  In an axially-symmetric geometry, the net
polarization can be non-zero and will align either parallel or
perpendicular to the axis of symmetry.  We use the convention that
positive (negative) polarization designates a polarization oriented
parallel (perpendicular) to the axis of symmetry.  SN~2001el had a
well-defined polarization angle over most of the spectral features,
which indicates that the bulk of the ejecta obeyed a near axial
symmetry (in addition, SN~2001el showed an unusual high velocity
\ion{Ca}{2} IR triple feature with a distinct polarization angle,
corresponding to a detached ``clump'' of material that deviated from
the dominant axis of symmetry\citep{Kasen_01el}).

Light becomes polarized in supernova atmospheres due to electron
scattering; other sources of opacity, such as bound-bound line
transitions, are usually considered to be depolarizing.  We define the
continuum polarization as the polarization computed using only
electron scattering opacity -- this is most closely realized in the
red end of a supernova spectrum (say near 7000~\AA), where there is
not much line opacity.  However this may not be the maximum
polarization level in the spectrum, as line opacity may partially
obscure the underlying photosphere and lead to a less effective
cancellation of the polarization in the line features (see
\S~\ref{polspec_section} and \cite{Kasen_01el}).  \cite{Hoeflich_91}
computes the continuum polarization in ellipsoidal and other
axially symmetric geometries.

Figure~\ref{pol_cont} shows the continuum polarization of the
ejecta-hole model as a function of viewing angle.  When viewed
directly down the hole ($\theta=0^\circ$) the projection of the
supernova atmosphere is circularly symmetric and the polarization
cancels.  As the viewing angle is inclined, the polarization
increases, reaching a maximum when the supernova is viewed nearly
side-on ($\theta \approx 90^\circ$).  The origin of the non-zero
polarization is clear from Figure~\ref{block_plot}a.  At inclinations
near $90^\circ$, the hole removes a ``wedge'' of scatterers from the
top of the atmosphere, which decreases the horizontally polarized flux
coming from this region.  The vertically polarized flux thus exceeds
the horizontal; the net polarization is non-zero and aligned with the
symmetry axis of the system (positive according to our convention).

To determine the level of intrinsic continuum polarization in observed
supernova, one must wrestle with the issue of subtracting the
interstellar polarization \citep{Howell_99by, Leonard_98s}.  Once this
is done, the observed levels are found to be rather small: the
polarization of SN~2001el was $\sim$0.3\%; the polarization of the
subluminous SN~1999by $\sim 0.7\%$.  For several other SNe~Ia, no
polarization signal was detected, but upper limits of 0.3-0.5\% can be
derived \citep{wang-broadband, wang_pol_96}.  In the ejecta-hole
model, the continuum polarization can be as large as 0.8\%, while the
polarization at the line features can be even larger (see next
section).  The hole asymmetry therefore produces polarization levels
in the right range, though perhaps generally too high compared to the
current published observations.  

The polarization in the ejecta-hole model, however, is rather
sensitive to the size and density of the hole.  To demonstrate this we
have over-plotted the continuum polarization of a model with a smaller
opening angle ($\theta_H = 30^\circ$).  This tames the asymmetry and
decrease the continuum polarization by more than a factor of two.  If
the hole size is decreased further to $\theta_H < 20^\circ$, the
continuum polarization level is uninterestingly small ($\la 0.1\%$)
from all inclinations.  Thus the exact polarization level will depend
upon the hole structure, which in turn depends upon the details of the
progenitor system and hydrodynamics.  In general, the more extreme the
asymmetry of the hole (i.e. the larger and more evacuated it is) the
higher the average polarization level.  A larger sample of SNe~Ia
spectropolarimetry could therefore put constraints on the size of a
putative hole.  Current observations may already constrain the hole to
have $\theta_H \la 40^\circ$.

One correlation to keep in mind is that the continuum polarization is
always relatively small ($\la 0.1\%$) for views near the hole where
the spectrum looks peculiar.  For views away from the hole, the
continuum polarization may be either small or large.  However the
continuum polarization is not the whole story and as we see in the
next section, the polarization over the line features can be
substantial even for $\theta < \theta_H$.

\subsection{Polarization Spectrum}
\label{polspec_section}

\begin{figure}
\begin{center}
\psfig{file=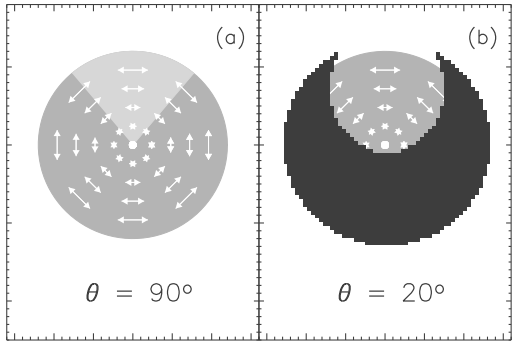,width=3.3in}
\caption{Schematic diagram which helps explain the polarization in the
ejecta hole model.  (a) When viewed from the side ($\theta \approx
90^\circ$) the top of the atmosphere is lacking a wedge of scatterers.
The vertically polarized flux thus exceeds the horizontal and the
continuum polarization is positive. (b) When viewed just off the
hole axis ($\theta \approx 20^\circ$), the line opacity on the planar
surface corresponding to a certain line of sight blueshift (shown in
black) only partially covers the photosphere.  Because of the hole,
horizontally polarized flux from the top of the atmosphere is
relatively unobscured by the line and will cause the negatively
polarized line peaks.
\label{block_plot}}
\end{center}
\end{figure}

\begin{figure}
\begin{center}
\psfig{file=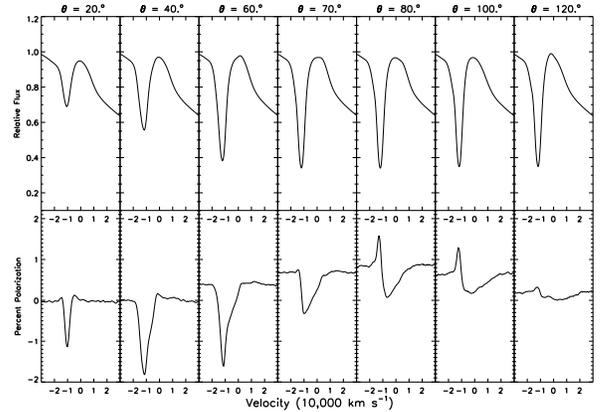,width=3.3in}
\caption{Flux and polarization profiles of a single unblended line in
the ejecta-hole geometry from various viewing angle. The polarization
feature is a negative (i.e. horizontally polarized) peak for
$(10^\circ < \theta < 60^\circ$) and an inverted P-Cygni for ($\theta
> 80^\circ$).
\label{line_pol}}
\end{center}
\end{figure}

The continuum polarization level actually provides very little
information as to the nature of the ejecta asymmetry, as very
different configurations can give the same numerical value.  Line
features in the polarization spectrum, on the other hand, contain more
potential information about the specific geometry.  We find that the
ejecta-hole model has spectropolarimetric signatures that distinguish
it from, for example, an ellipsoidal geometry.

The polarization spectrum in the ellipsoidal geometry has been studied
in detail \citep{Jeffery-Sobolev-P, Hoflich-93J}.  In ellipsoidal
models, the polarization level generally increases from blue to red
due to the greater amount of depolarizing line opacity in the blue.
Individual lines create ``inverted P-Cygni'' profiles in the
polarization spectrum, i.e. a blueshifted polarization peak with a
redshifted depolarization trough.  The blueshifted peak is a result of
the line opacity preferentially blocking the lowly polarized central
photospheric light, while the redshifted trough is the result of
unpolarized line emission light diluting the continuum polarization.
The polarized line profiles look fairly similar from all viewing
angles.

The line polarization profile in the ejecta-hole model shows an
interesting variation with inclination (Figure~\ref{line_pol}).  For
views far enough away from the hole ($\theta \ga 80^\circ$), the
profile is an inverted P-Cygni, just as in an ellipsoidal model, and
for essentially the same reason.  For views closer to the hole,
however, the blueshifted line absorption gives rise to a large
polarization peak (recall the negative sign indicates that the
polarization direction is perpendicular to the symmetry axis of the
system).  Figure~\ref{block_plot}b helps explain the origin of the
peak.  From viewing angles near the hole axis, the projected electron
scattering medium is fairly symmetric and the continuum polarization
integrated over the ejecta surface nearly cancels.  The line opacity,
however, only partially obscures the underlying light.  Because of the
hole, horizontally polarized flux from the top of the atmosphere is
relatively unobscured, whereas the vertically polarized light from
the sides of the atmosphere is effectively screened by the line.  The
polarization over the line therefore does not cancel, but will be
large and oriented perpendicular to the axis of symmetry (negative
according to our convention).  Note that if the hole opening angle is
narrowed to $\theta = 30^\circ$, the line is even more effective in
screening off all but the horizontally polarized light.  The line
polarization peak is therefore \emph{larger}.  Thus while the
continuum polarization decreases with decreasing hole size, the line
polarization from certain viewing angles will be relatively large
($\ga 1.0 \%$) regardless of how big the hole is.

\begin{figure}
\begin{center}
\psfig{file=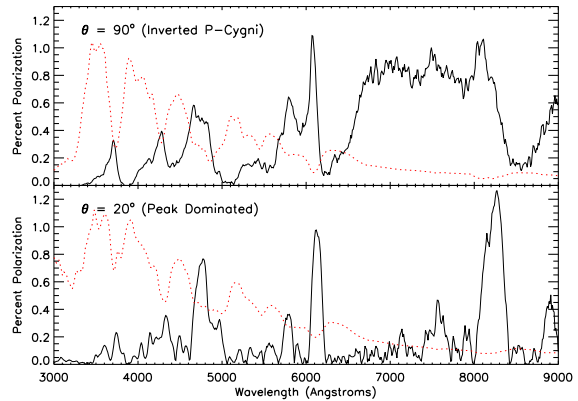,width=3.3in}
\caption{Polarization spectrum of the ejecta-hole model near maximum
light for two viewing angles.  We plot the absolute value of the
polarization (solid lines), and for reference overlay the flux
spectrum (dotted lines).  The small scale wiggles in the polarization
spectrum are Monte Carlo noise, which increases to the red due to the
lower fluxes. Top: for views away from the hole (here $\theta =
90^\circ$) the spectrum resembles that of an ellipsoidal geometry with
``inverted P-Cygni'' line profiles.  Bottom: for views nearer the hole
(here $\theta = 20^\circ$), the spectrum is ``peak-dominated'' with a
low continuum polarization but substantial line peaks.
\label{pol_spec}}
\end{center}
\end{figure}

Figure~\ref{pol_spec} shows the entire ejecta-hole polarization
spectrum from two lines of sight.  For a view near the hole ($\theta =
20^\circ$) the spectrum is ``line peak-dominated'' -- the continuum
polarization is rather low, but large polarization peaks are
associated with the blueshifted line absorption features (in
particular the \ion{Si}{2} 6150 feature and the \ion{Ca}{2} IR
triplet).  This spectrum is qualitatively different than what is
expected in an ellipsoidal geometry.  For views away from the hole
($\theta = 90^\circ)$, on the other hand, the polarization spectrum
would be very hard to distinguish from the ellipsoidal case.  The
level of polarization rises from blue to red and the line features due
to \ion{Si}{2} 6150 feature and \ion{Ca}{2} IR triplet have the
``inverted P-Cygni'' profile.  The shape of the polarization spectrum
from these angles resembles SN~2001el, although the polarization level
is too high unless $\theta \ga 110^\circ$, or the hole opening angle
is reduced.

To discriminate between different geometries, a larger sample of
polarization spectra is needed.  If the asymmetry in SNe~Ia is an
ejecta hole, we would expect to see something like a line peak
dominated polarization spectrum for $10^\circ \la \theta \la
60^\circ$, or about 25\% of the time.  Such a polarization spectrum
has not been observed as yet, but the number of published
spectropolarimetric observations is still relatively small.
Uncertainty in the interstellar polarization may make it difficult to
identify the peaks, for if the zeropoint of the intrinsic supernova
polarization is unknown, it will be unclear whether features in the
polarization spectrum are peaks or troughs.  Therefore multi-epoch
spectropolarimetric observations are necessary to help pin down the
interstellar component.  Of course, observing a line peak dominated
polarization spectrum may not uniquely implicate an ejecta hole, as
large line peaks could potentially occur in other geometries so far
unexplored.

\section{Conclusions}

\subsection{Asymmetry and Diversity in SNe~Ia}

Despite the seemingly extreme nature of an ejecta-hole asymmetry, we
find that the geometry is actually consistent with what is currently
known about SNe~Ia, at least for the observables we have calculated.
The variation of the peak magnitude with viewing angle is $\sim 0.2$
mag in $B$, comparable to the intrinsic dispersion of SNe~Ia, and the
level of polarization in the range observed ($0-0.8\%$).  The spectrum
of the supernova looks peculiar when viewed near the hole, but this
peculiarity may fit in with the spectral diversity already known to
exist in SNe~Ia. In addition, the polarization spectrum from some
lines of sight is a qualitative match to that of SN~2001el.

An ejecta-hole asymmetry could therefore be one source of diversity in
SNe~Ia, but of course not the only one.  The primary source of
diversity in SNe~Ia is thought to be due to variations in the amount
of \Ni produced in the explosion.  Because SNe~Ia are powered by the
radioactive decay of \Ni and its daughter $^{56}$Co, different \Ni
production can explain the variety in SN~Ia peak magnitudes.  Greater
\Ni masses may lead to higher atmospheric temperatures and higher
effective opacities, which may explain why brighter SNe~Ia have
broader light curves
\citep{Hoeflich_WLR,Pinto_Eastman_I,Nugent_hydro}.

It has often been thought that the spectroscopic diversity of SNe~Ia
fits into the same one-parameter \Ni sequence \citep{Nugent_sequence}.
In this picture, SN~1991T-like supernovae occupy the overluminous end
of the sequence, where the larger \Ni mass leads to higher envelope
temperatures and a higher ionization fraction.  This may explain the
peculiar spectral appearance \citep{Mazzali_91T,Jeffery_92}.  As the
models in this paper show, there could be a second, physically very
different route to the same sort of spectral peculiarities -- one
could be peering down an ejecta hole.  In the ejecta-hole model, the
spectrum shows some level of peculiarity for $\theta \la \theta_H$ or
about 12\% of the time, although the peculiarities will only be very
intense for views more directly down the hole ($\theta \la \theta_H/2$
or $\sim 3\%$ of the time).  The observed rate of SN~1991T-like
supernovae is $\sim 3-5$\% in the samples of both
\cite{Branch_peculiar} and \cite{Li_peculiar}; the rate of
SN~1991T/SN~1999aa-like supernovae is $20\% \pm 7\%$ in the
\cite{Li_peculiar} sample.  Therefore it is possible that a
substantial percentage of these peculiar supernovae could be the
result of an ejecta-hole asymmetry.

In this paper we have chosen to compare the spectra emanating from the
hole with SN~1991T only because it is the well-known prototype of a
certain kind of spectral peculiarity.  Whether SN~1991T itself was an
example of merely looking down an ejecta hole is debatable.  Initial
estimates suggested that SN~1991T was as much as 0.7-0.8 magnitudes
brighter in $B$ than normal, which is too much to be explained by the
asymmetry alone \citep{Fisher_91T}.  More recent Cepheid measurements
of the distance to the host galaxy, however, show that SN~1991T was
not really much brighter than a normal SN~Ia.  \citep{Saha_91T} find a
moderate overluminosity of 0.3 mag, although a value as high 0.6 mag
cannot be ruled out due to large uncertainty in the dust extinction.
This lower value for the brightness of SN~1991T calls into question
whether the peculiar spectral appearance can still be explained alone
by high envelope temperatures due to a larger \Ni mass.

SN~1991T also had a rather broad light curve (\dmf = 0.95 $\pm$ 0.05;
\cite{Phillips99}), which is often taken as an indication of a large
\Ni mass.  Because we have not yet computed time-dependent models, we
do not know exactly what effect an ejecta hole asymmetry will have on
the light curve.  Because the hole acts as an energy leak, it probably
lowers the net diffusion time, and we expect that the
\emph{integrated} light curve (i.e that summed over all viewing
angles) will be narrower in a ejecta-hole model than a spherical
model.  But the real question relevant to SN~1991T is not how the
integrated light curve compares to a spherical model, but whether the
light curve viewed down the hole is broader or narrower than that from
other viewing angles.  In other words we need to know how
Figure~\ref{mags} -- the distribution of the total luminosity among
viewing angles -- varies with time.  This is more difficult to intuit,
because as the ejecta thins out and the asymmetry and opacities evolve
with time, it is hard to say off-hand whether it will become more or
less easy for photons to preferentially escape out the hole.  We leave
the question for future work.

In any case, although the prototype SN~1991T did have a broad light
curve, it is not clear whether a general correlation between light curve
width and SN~1991T-like spectral peculiarities even exists
\citep{Howell_01ay}.  Several SNe~Ia have similar or broader light
curves, and yet the spectrum is apparently normal -- at least eight
such supernovae with \dmf $< 1.0$ are listed in \cite{Phillips99}, for
example SN~1992bc (\dmf = 0.87 $\pm$ 0.05) and SN~1994ae (\dmf = 0.86
$\pm$ 0.05).  SN~2001ay also had a normal spectrum but an
exceptionally broad light curve (\dmf = 0.6-0.7;
\cite{Phillips_01ay}).  Among the supernovae with SN~1991T-like
spectral peculiarities, there also appears to be diversity.  SN~1997br
had a moderately broad light curve (\dmf $= 1.00 \pm 0.15$;
\cite{Li_97br}), but the light curve of SN~2002cx was on the narrow
side (\dmf $= 1.30 \pm 0.09$; \cite{Li_02cx}).  In another
SN~1991T-like supernova the $B$-band light curve was lopsided --
SN~2000cx brightened much faster than SN~1991T (resembling the rise of
the normal SN~1994D) but the decline was slow (\dmf = 0.93 $\pm$ 0.04;
\cite{Li_00cx}).  The examples make it clear that the connection
between light-curve width and SN~1991T-like spectral peculiarities
remains vague, and that more than one parameter of diversity needs to
be identified.

The nebular spectra of SN~1991T may also suggest a large \Ni
production.  In the late time spectra, the iron emission lines of
SN~1991T have larger velocity widths than in most
SNe~Ia \citep{Mazzali_nebular}.  Assuming the late time
ionization/excitation conditions are similar in all SNe~Ia, this
implies that the nickel/iron core in SN~1991T is larger than normal.
Confusing this conclusion, however, is the fact pointed out by
\cite{Hatano_97br} that the \ion{Si}{2} velocities in the post maximum
spectra are among the \emph{lowest} of all SNe~Ia.  If SN~1991T really
did have a large inner \Ni zone, one naively expects the zone of
silicon and other intermediate mass elements to occur at especially
high velocities (as for instance in the delayed detonation models of
e.g. \cite{Hoeflich_99by}).  To account for the low \ion{Si}{2}
velocities, some have invoked a late-detonation model for SN~1991T,
which produces a layer of intermediate mass elements sandwiched
between two nickel zones \citep{Yamaoka_91T,Pilar_91T}.  Of course a
lower \ion{Si}{2} velocity is also naturally expected if one is
looking down an ejecta hole.

It is possible that SN~1991T did have a relatively large \Ni mass,
rather than (or perhaps in addition to) being viewed down the ejecta
hole.  However among other supernovae with SN~1991T-like peculiarities
there is a good deal of diversity, and the large \Ni mass explanation
will not apply in all cases.  The most obvious case in point is
SN~2002cx \citep{Li_02cx}.  The spectrum of SN~2002cx resembled
SN~1991T in that \ion{Si}{2}, \ion{S}{2}, and \ion{Ca}{2} lines were
weak, while \ion{Fe}{3} was prominent, but the supernova was
\emph{underluminous} by $\sim2$ mag.  The velocities of the absorption
features were also unusually low ($v \approx 7.000$ \kms)
\citep{Branch_02cx}.  The singularity of the supernova led
\cite{Li_02cx} to consider alternative progenitor systems, although
they conclude that no existing theoretical model could explain all the
peculiarities.  On the other hand, there is also the possibility that
we are seeing multiple channels of diversity operating at once -- one
scenario to entertain now is that we are looking down the ejecta hole
of a ``weak'' supernova that produced a small mass of $^{56}$Ni.  Such
underluminous objects (e.g. SN~1991bg) typically have relatively low
absorption velocities \citep{Turatto_91bg,Mazzali_91bg,Hatano_97br}
which would be further reduced by looking down the hole.  Despite the
low luminosity, the spectrum might still appear hot and iron dominated
if one is peering into the deeper layers, and (eventually) into the
iron core.  Of course, the chance of seeing two distinct sources of
diversity operating at once would be, like SN~2002cx, a very rare
occurrence.

Whatever the final explanation for SN~2002cx, its singularity
highlights the fact that the diversity of SNe~Ia is more complicated
than a one-parameter sequence based upon $^{56}$Ni.  One can imagine
many sources of variation contributing, including: different amounts
and distributions of high-velocity material, different sizes and
inclinations of an ejecta hole (or other) asymmetry, various explosion
processes (e.g. pure deflagration, deflagration-to-detonation,
late-detonation), metallicity variations and even the possibility of
completely different progenitor channels (e.g. a double-degenerate
scenario).  With so much going on at once, interpreting the
observations becomes a daunting task, but in some sense the
identification of multiple channels of diversity is welcome,
considering that observations continue to turn up unusual SNe~Ia.

\subsection{Observational Consequence of an Ejecta Hole}

The results of this paper suggest a few observational signatures of
the ejecta-hole geometry.  First, the continuum polarization should be
low for views directly down the hole, where the spectrum looks
peculiar.  However because of the partial obscuration effect, the
polarization spectrum should show large line peaks for views just away
the hole ($10^\circ < \theta < 60^\circ$), where the spectrum looks
marginally peculiar or normal.  For views from the side ($\theta
\approx 90^\circ$), a relatively high continuum polarization should be
correlated with a slightly dimmer supernova with normal spectral
features and inverted P-Cygni line polarization features.  Another
possible signature of the ejecta hole is ``lopsided'' P-Cygni flux
profiles -- the view down the hole weakens only the absorption, not
the emission feature, so one could look for a weak (or absent)
absorption associated with noticeable emission.  The easiest place to
look would be in the \ion{Si}{2} 6150 and the \ion{Ca}{2} IR triplet
features of SN~1991T-like supernova.  Unfortunately the relative
strength of absorption to emission depends also on the line source
function, which is determined by the detailed excitation conditions in
the atmosphere.  In general, because we recognize that an ejecta-hole
asymmetry is only one of several possible sources of diversity in
SNe~Ia, it will be difficult to isolate the geometrical effects from
the other variations that may be operating.  The only hope is to
collect a large sample of supernovae with well observed light curves,
spectra and polarization, so that one might try to pull out the
different trends.

In our calculations we have used a parameterized hole (half opening
angle $40^\circ$) in order to explore the essential observable
consequences of the geometry.  The next step is to address the same
questions using specific hydrodynamical models representing a wide
variety of progenitor configurations.  The details of the progenitor
system could potentially affect the size and shape of the hole.
\cite{Marietta} compute interactions using main sequence, subgiant and
red giant companions and note that the variation in the hole asymmetry
is not large.  This is because in all cases the companion star is near
enough to have undergone Roche lobe overflow and always occupies a
similar solid angle (the red giant is farther away but physically
larger than a main sequence companion which is smaller but much
closer).  However if the ratio of companion radius to separation
distance is decreased for some reason, the size of the hole also
decreases.  A larger sample of spectropolarimetric observations will
help determine if SNe~Ia really do have an ejecta-hole geometry, and
could constrain the hole opening angle if one exists.  While a hole
smaller than $\theta < 20^\circ$ has only minor effects on the
spectrum, luminosity and continuum polarization, it will still create
substantial line peaks in the polarization spectrum when seen from
some viewing angles.  If such signatures of the hole are not seen in
future spectropolarimetric observations, this would have interesting
consequences for the progenitors of SNe~Ia, or the hydrodynamics of
the ejecta/companion interaction.

Finally, we mention that an asymmetry like an ejecta hole could have a
number of subtle consequences on the use of SNe~Ia as standard candles
for cosmology.  The asymmetry causes a $\sim 20\%$ dispersion in
observed SNe~Ia peak magnitude.  If the asymmetry is identical in all
supernova, this dispersion behaves like a statistical error (although a
non-gaussian one) and can be averaged out by observing enough objects.
The averaging out is not achieved, however, if one does not
sufficiently sample \emph{every} possible viewing angle, either
because not enough supernovae are observed, or because those viewed
down the hole are withheld from the sample due to concern over their
spectral peculiarities.  In addition, if the nature, degree, or
frequency of the asymmetry evolves with redshift (say because of
evolving progenitor populations) the peak magnitude of SNe~Ia becomes
a function of redshift.  One might also be concerned that the
significant angular variation of the colors and spectrum may
complicate extinction and K-corrections. The errors incurred from all
these and other related systematic effects would be relatively small,
but may need to be considered in the next generation of precision
supernova cosmology experiments.


\begin{thebibliography}{60}
\expandafter\ifx\csname natexlab\endcsname\relax\def\natexlab#1{#1}\fi

\bibitem[{{Arnett}(1982)}]{Arnett_typeI}
{Arnett}, W.~D. 1982, \apj, 253, 785

\bibitem[{{Branch}(1987)}]{Branch_84A}
{Branch}, D. 1987, \apjl, 316, L81

\bibitem[{{Branch}(2001)}]{Branch_peculiar}
---. 2001, \pasp, 113, 169

\bibitem[{{Branch}(2003)}]{Branch_02cx}
---. 2003, To appear in 3-D Signatures of Stellar Explosions: astro-ph/0310685

\bibitem[{{Branch} {et~al.}(1985){Branch}, {Doggett}, {Nomoto}, \&
  {Thielemann}}]{Branch_81b}
{Branch}, D., {Doggett}, J.~B., {Nomoto}, K., \& {Thielemann}, F.-K. 1985,
  \apj, 294, 619

\bibitem[{{Branch} {et~al.}(1993){Branch}, {Fisher}, \&
  {Nugent}}]{Branch_normal}
{Branch}, D., {Fisher}, A., \& {Nugent}, P. 1993, \aj, 106, 2383

\bibitem[{{Branch} {et~al.}(1995){Branch}, {Livio}, {Yungelson}, {Boffi}, \&
  {Baron}}]{Branch_progenitor}
{Branch}, D., {Livio}, M., {Yungelson}, L.~R., {Boffi}, F.~R., \& {Baron}, E.
  1995, \pasp, 107, 1019

\bibitem[{{Branch} {et~al.}(2003)}]{Branch_00cx}
{Branch}, D. {et~al.} 2003, ApJ, in preparation

\bibitem[{{Code} \& {Whitney}(1995)}]{Code-blobs}
{Code}, A.~D. \& {Whitney}, B.~A. 1995, \apj, 441, 400

\bibitem[{{Filippenko} {et~al.}(1992{\natexlab{a}}){Filippenko}, {Richmond},
  {Branch}, {Gaskell}, {Herbst}, {Ford}, {Treffers}, {Matheson}, {Ho}, {Dey},
  {Sargent}, {Small}, \& {van Breugel}}]{Filippenko_91bg}
{Filippenko}, A.~V., {Richmond}, M.~W., {Branch}, D., {Gaskell}, M., {Herbst},
  W., {Ford}, C.~H., {Treffers}, R.~R., {Matheson}, T., {Ho}, L.~C., {Dey}, A.,
  {Sargent}, W.~L.~W., {Small}, T.~A., \& {van Breugel}, W.~J.~M.
  1992{\natexlab{a}}, \aj, 104, 1543

\bibitem[{{Filippenko} {et~al.}(1992{\natexlab{b}}){Filippenko}, {Richmond},
  {Matheson}, {Shields}, {Burbidge}, {Cohen}, {Dickinson}, {Malkan}, {Nelson},
  {Pietz}, {Schlegel}, {Schmeer}, {Spinrad}, {Steidel}, {Tran}, \&
  {Wren}}]{Filippenko_91T}
{Filippenko}, A.~V., {Richmond}, M.~W., {Matheson}, T., {Shields}, J.~C.,
  {Burbidge}, E.~M., {Cohen}, R.~D., {Dickinson}, M., {Malkan}, M.~A.,
  {Nelson}, B., {Pietz}, J., {Schlegel}, D., {Schmeer}, P., {Spinrad}, H.,
  {Steidel}, C.~C., {Tran}, H.~D., \& {Wren}, W. 1992{\natexlab{b}}, \apjl,
  384, L15

\bibitem[{{Fisher} {et~al.}(1999){Fisher}, {Branch}, {Hatano}, \&
  {Baron}}]{Fisher_91T}
{Fisher}, A., {Branch}, D., {Hatano}, K., \& {Baron}, E. 1999, \mnras, 304, 67

\bibitem[{{Fryxell} \& {Arnett}(1981)}]{Fryxell_Arnett}
{Fryxell}, B.~A. \& {Arnett}, W.~D. 1981, \apj, 243, 994

\bibitem[{{H{\" o}flich} {et~al.}(2002){H{\" o}flich}, {Gerardy}, {Fesen}, \&
  {Sakai}}]{Hoeflich_99by}
{H{\" o}flich}, P., {Gerardy}, C.~L., {Fesen}, R.~A., \& {Sakai}, S. 2002,
  \apj, 568, 791

\bibitem[{{Hamuy} {et~al.}(1996){Hamuy}, {Phillips}, {Suntzeff}, {Schommer},
  {Maza}, \& {Aviles}}]{Hamuy_96b}
{Hamuy}, M., {Phillips}, M.~M., {Suntzeff}, N.~B., {Schommer}, R.~A., {Maza},
  J., \& {Aviles}, R. 1996, \aj, 112, 2391

\bibitem[{{Hatano} {et~al.}(1999){Hatano}, {Branch}, {Fisher}, {Baron}, \&
  {Filippenko}}]{Hatano_94d}
{Hatano}, K., {Branch}, D., {Fisher}, A., {Baron}, E., \& {Filippenko}, A.~V.
  1999, ApJ, 525, 881

\bibitem[{{Hatano} {et~al.}(2002){Hatano}, {Branch}, {Qiu}, {Baron},
  {Thielemann}, \& {Fisher}}]{Hatano_97br}
{Hatano}, K., {Branch}, D., {Qiu}, Y.~L., {Baron}, E., {Thielemann}, F.-K., \&
  {Fisher}, A. 2002, New Astronomy, 7, 441

\bibitem[{{H{\"o}flich}(1991)}]{Hoeflich_91}
{H{\"o}flich}, P. 1991, \aap, 246, 481+

\bibitem[{{Hoflich} {et~al.}(1995){Hoflich}, {Khokhlov}, \&
  {Wheeler}}]{Hoeflich_WLR}
{Hoflich}, P., {Khokhlov}, A.~M., \& {Wheeler}, J.~C. 1995, \apj, 444, 831

\bibitem[{{H{\"o}flich} {et~al.}(1996){H{\"o}flich}, {Wheeler}, {Hines}, \&
  {Trammell}}]{Hoflich-93J}
{H{\"o}flich}, P., {Wheeler}, J.~C., {Hines}, D.~C., \& {Trammell}, S.~R. 1996,
  \apj, 459, 307+

\bibitem[{{Howell}(2003)}]{Howell_01ay}
{Howell}, D. 2003, To appear in 3-D Signatures of Stellar Explosions

\bibitem[{{Howell} {et~al.}(2001){Howell}, {H{\" o}flich}, {Wang}, \&
  {Wheeler}}]{Howell_99by}
{Howell}, D.~A., {H{\" o}flich}, P., {Wang}, L., \& {Wheeler}, J.~C. 2001,
  \apj, 556, 302

\bibitem[{{Jeffery} {et~al.}(1992){Jeffery}, {Leibundgut}, {Kirshner},
  {Benetti}, {Branch}, \& {Sonneborn}}]{Jeffery_92}
{Jeffery}, D.~J., {Leibundgut}, B., {Kirshner}, R.~P., {Benetti}, S., {Branch},
  D., \& {Sonneborn}, G. 1992, \apj, 397, 304

\bibitem[{{Jeffrey}(1989)}]{Jeffery-Sobolev-P}
{Jeffrey}, D.~J. 1989, \apjs, 71, 951

\bibitem[{{Jeffrey}(1991)}]{Jeffery_87a}
---. 1991, \apj, 375, 264

\bibitem[{{Kasen} {et~al.}(2003{\natexlab{a}}){Kasen}, {Nugent}, {Wang},
  {Howell}, {Wheeler}, {H{\" o}flich}, {Baade}, {Baron}, \&
  {Hauschildt}}]{Kasen_01el}
{Kasen}, D., {Nugent}, P., {Wang}, L., {Howell}, D.~A., {Wheeler}, J.~C., {H{\"
  o}flich}, P., {Baade}, D., {Baron}, E., \& {Hauschildt}, P.~H.
  2003{\natexlab{a}}, \apj, 593, 788

\bibitem[{{Kasen} {et~al.}(2003{\natexlab{b}})}]{Kasen_MC}
{Kasen}, D. {et~al.} 2003{\natexlab{b}}, in preparation

\bibitem[{{Lentz} {et~al.}(2001){Lentz}, {Baron}, {Branch}, \&
  {Hauschildt}}]{Lentz_94d}
{Lentz}, E.~J., {Baron}, E., {Branch}, D., \& {Hauschildt}, P.~H. 2001, \apj,
  557, 266

\bibitem[{{Leonard} {et~al.}(2000){Leonard}, {Filippenko}, {Barth}, \&
  {Matheson}}]{Leonard_98s}
{Leonard}, D.~C., {Filippenko}, A.~V., {Barth}, A.~J., \& {Matheson}, T. 2000,
  \apj, 536, 239

\bibitem[{{Li} {et~al.}(2003){Li}, {Filippenko}, {Chornock}, {Berger},
  {Berlind}, {Calkins}, {Challis}, {Fassnacht}, {Jha}, {Kirshner}, {Matheson},
  {Sargent}, {Simcoe}, {Smith}, \& {Squires}}]{Li_02cx}
{Li}, W., {Filippenko}, A.~V., {Chornock}, R., {Berger}, E., {Berlind}, P.,
  {Calkins}, M.~L., {Challis}, P., {Fassnacht}, C., {Jha}, S., {Kirshner},
  R.~P., {Matheson}, T., {Sargent}, W.~L.~W., {Simcoe}, R.~A., {Smith}, G.~H.,
  \& {Squires}, G. 2003, \pasp, 115, 453

\bibitem[{{Li} {et~al.}(2001{\natexlab{a}}){Li}, {Filippenko}, {Gates},
  {Chornock}, {Gal-Yam}, {Ofek}, {Leonard}, {Modjaz}, {Rich}, {Riess}, \&
  {Treffers}}]{Li_00cx}
{Li}, W., {Filippenko}, A.~V., {Gates}, E., {Chornock}, R., {Gal-Yam}, A.,
  {Ofek}, E.~O., {Leonard}, D.~C., {Modjaz}, M., {Rich}, R.~M., {Riess}, A.~G.,
  \& {Treffers}, R.~R. 2001{\natexlab{a}}, \pasp, 113, 1178

\bibitem[{{Li} {et~al.}(2001{\natexlab{b}}){Li}, {Filippenko}, {Treffers},
  {Riess}, {Hu}, \& {Qiu}}]{Li_peculiar}
{Li}, W., {Filippenko}, A.~V., {Treffers}, R.~R., {Riess}, A.~G., {Hu}, J., \&
  {Qiu}, Y. 2001{\natexlab{b}}, \apj, 546, 734

\bibitem[{{Li} {et~al.}(1999){Li}, {Qiu}, {Qiao}, {Zhu}, {Hu}, {Richmond},
  {Filippenko}, {Treffers}, {Peng}, \& {Leonard}}]{Li_97br}
{Li}, W.~D., {Qiu}, Y.~L., {Qiao}, Q.~Y., {Zhu}, X.~H., {Hu}, J.~Y.,
  {Richmond}, M.~W., {Filippenko}, A.~V., {Treffers}, R.~R., {Peng}, C.~Y., \&
  {Leonard}, D.~C. 1999, \aj, 117, 2709

\bibitem[{{Livne} {et~al.}(1992){Livne}, {Tuchman}, \&
  {Wheeler}}]{Livne_binary}
{Livne}, E., {Tuchman}, Y., \& {Wheeler}, J.~C. 1992, \apj, 399, 665

\bibitem[{{Lucy}(1999)}]{Lucy_Radeq}
{Lucy}, L.~B. 1999, \aap, 344, 282

\bibitem[{{Marietta} {et~al.}(2000){Marietta}, {Burrows}, \&
  {Fryxell}}]{Marietta}
{Marietta}, E., {Burrows}, A., \& {Fryxell}, B. 2000, \apjs, 128, 615

\bibitem[{{Mazzali} {et~al.}(1998){Mazzali}, {Cappellaro}, {Danziger},
  {Turatto}, \& {Benetti}}]{Mazzali_nebular}
{Mazzali}, P.~A., {Cappellaro}, E., {Danziger}, I.~J., {Turatto}, M., \&
  {Benetti}, S. 1998, \apjl, 499, L49+

\bibitem[{{Mazzali} {et~al.}(1997){Mazzali}, {Chugai}, {Turatto}, {Lucy},
  {Danziger}, {Cappellaro}, {della Valle}, \& {Benetti}}]{Mazzali_91bg}
{Mazzali}, P.~A., {Chugai}, N., {Turatto}, M., {Lucy}, L.~B., {Danziger},
  I.~J., {Cappellaro}, E., {della Valle}, M., \& {Benetti}, S. 1997, \mnras,
  284, 151

\bibitem[{{Mazzali} {et~al.}(1995){Mazzali}, {Danziger}, \&
  {Turatto}}]{Mazzali_91T}
{Mazzali}, P.~A., {Danziger}, I.~J., \& {Turatto}, M. 1995, \aap, 297, 509

\bibitem[{{Mazzali} \& {Lucy}(1993)}]{Mazzali_MC}
{Mazzali}, P.~A. \& {Lucy}, L.~B. 1993, \aap, 279, 447

\bibitem[{Nomoto {et~al.}(1984)Nomoto, Thielemann, \& Yokoi}]{Nomoto_w7}
Nomoto, K., Thielemann, F., \& Yokoi, K. 1984, ApJ, 286, 644

\bibitem[{{Nugent} {et~al.}(1997){Nugent}, {Baron}, {Branch}, {Fisher}, \&
  {Hauschildt}}]{Nugent_hydro}
{Nugent}, P., {Baron}, E., {Branch}, D., {Fisher}, A., \& {Hauschildt}, P.~H.
  1997, \apj, 485, 812

\bibitem[{{Nugent} {et~al.}(1995){Nugent}, {Phillips}, {Baron}, {Branch}, \&
  {Hauschildt}}]{Nugent_sequence}
{Nugent}, P., {Phillips}, M., {Baron}, E., {Branch}, D., \& {Hauschildt}, P.
  1995, \apjl, 455, L147+

\bibitem[{{Phillips}(1993)}]{Phillips93}
{Phillips}, M.~M. 1993, \apjl, 413, L105

\bibitem[{{Phillips} {et~al.}(2003){Phillips}, {Krisciunas}, {Suntzeff},
  {Roth}, {Germany}, {Candia}, {Gonzalez}, {Hamuy}, {Freedman}, {Persson},
  {Nugent}, {Aldering}, \& {Conley}}]{Phillips_01ay}
{Phillips}, M.~M., {Krisciunas}, K., {Suntzeff}, N.~B., {Roth}, M., {Germany},
  L., {Candia}, P., {Gonzalez}, S., {Hamuy}, M., {Freedman}, W.~L., {Persson},
  S.~E., {Nugent}, P.~E., {Aldering}, G., \& {Conley}, A. 2003, in From
  Twilight to Highlight: The Physics of Supernovae. Proceedings of the
  ESO/MPA/MPE Workshop held in Garching, Germany, 29-31 July 2002, p. 193.,
  193--+

\bibitem[{{Phillips} {et~al.}(1999){Phillips}, {Lira}, {Suntzeff}, {Schommer},
  {Hamuy}, \& {Maza}}]{Phillips99}
{Phillips}, M.~M., {Lira}, P., {Suntzeff}, N.~B., {Schommer}, R.~A., {Hamuy},
  M., \& {Maza}, J. 1999, \aj, 118, 1766

\bibitem[{{Phillips} {et~al.}(1992){Phillips}, {Wells}, {Suntzeff}, {Hamuy},
  {Leibundgut}, {Kirshner}, \& {Foltz}}]{Phillips_91T}
{Phillips}, M.~M., {Wells}, L.~A., {Suntzeff}, N.~B., {Hamuy}, M.,
  {Leibundgut}, B., {Kirshner}, R.~P., \& {Foltz}, C.~B. 1992, \aj, 103, 1632

\bibitem[{{Pinto} \& {Eastman}(2000{\natexlab{a}})}]{Pinto_Eastman_I}
{Pinto}, P.~A. \& {Eastman}, R.~G. 2000{\natexlab{a}}, \apj, 530, 744

\bibitem[{{Pinto} \& {Eastman}(2000{\natexlab{b}})}]{Pinto-Eastman_II}
---. 2000{\natexlab{b}}, \apj, 530, 757

\bibitem[{{Ruiz-Lapuente} {et~al.}(1992){Ruiz-Lapuente}, {Cappellaro},
  {Turatto}, {Gouiffes}, {Danziger}, {della Valle}, \& {Lucy}}]{Pilar_91T}
{Ruiz-Lapuente}, P., {Cappellaro}, E., {Turatto}, M., {Gouiffes}, C.,
  {Danziger}, I.~J., {della Valle}, M., \& {Lucy}, L.~B. 1992, \apjl, 387, L33

\bibitem[{{Saha} {et~al.}(2001){Saha}, {Sandage}, {Thim}, {Labhardt},
  {Tammann}, {Christensen}, {Panagia}, \& {Macchetto}}]{Saha_91T}
{Saha}, A., {Sandage}, A., {Thim}, F., {Labhardt}, L., {Tammann}, G.~A.,
  {Christensen}, J., {Panagia}, N., \& {Macchetto}, F.~D. 2001, \apj, 551, 973

\bibitem[{{Thomas} {et~al.}(2002){Thomas}, {Kasen}, {Branch}, \&
  {Baron}}]{Rollin_clumpy}
{Thomas}, R.~C., {Kasen}, D., {Branch}, D., \& {Baron}, E. 2002, \apj, 567,
  1037

\bibitem[{{Thomas} {et~al.}(2003)}]{Thomas_00cx}
{Thomas}, R.~C. {et~al.} 2003, ApJ, submitted; astro-ph/0302260

\bibitem[{{Turatto} {et~al.}(1996){Turatto}, {Benetti}, {Cappellaro},
  {Danziger}, {della Valle}, {Gouiffes}, {Mazzali}, \& {Patat}}]{Turatto_91bg}
{Turatto}, M., {Benetti}, S., {Cappellaro}, E., {Danziger}, I.~J., {della
  Valle}, M., {Gouiffes}, C., {Mazzali}, P.~A., \& {Patat}, F. 1996, \mnras,
  283, 1

\bibitem[{{Wang} {et~al.}(1997){Wang}, {Wheeler}, \& {Hoeflich}}]{Wang_96x}
{Wang}, L., {Wheeler}, J.~C., \& {Hoeflich}, P. 1997, \apjl, 476, L27

\bibitem[{{Wang} {et~al.}(1996{\natexlab{a}}){Wang}, {Wheeler}, {Li}, \&
  {Clocchiatti}}]{wang-broadband}
{Wang}, L., {Wheeler}, J.~C., {Li}, Z., \& {Clocchiatti}, A.
  1996{\natexlab{a}}, \apj, 467, 435+

\bibitem[{{Wang} {et~al.}(1996{\natexlab{b}}){Wang}, {Wheeler}, {Li}, \&
  {Clocchiatti}}]{wang_pol_96}
---. 1996{\natexlab{b}}, ApJ, 467, 435

\bibitem[{Wang {et~al.}(2003)}]{Wang_pol_01el}
Wang, L. {et~al.} 2003, ApJ, in press

\bibitem[{{Wheeler} {et~al.}(1975){Wheeler}, {Lecar}, \&
  {McKee}}]{Wheeler_strip}
{Wheeler}, J.~C., {Lecar}, M., \& {McKee}, C.~F. 1975, \apj, 200, 145

\bibitem[{{Yamaoka} {et~al.}(1992){Yamaoka}, {Nomoto}, {Shigeyama}, \&
  {Thielemann}}]{Yamaoka_91T}
{Yamaoka}, H., {Nomoto}, K., {Shigeyama}, T., \& {Thielemann}, F. 1992, \apjl,
  393, L55

\end{thebibliography}

\end{document}